# Mobile Robots in Teaching Programming for IT Engineers and its Effects

Attila Pásztor, Róbert Pap-Szigeti, Erika Török
Kecskemét College, Faculty of Mechanical Engineering and Automation
Kecskemét, Hungary

*Abstract*—in this paper the new methods and devices introduced into the learning process of programming for IT engineers at our college is described. Based on our previous research results we supposed that project methods and some new devices can reduce programming problems during the first term. These problems are rooted in the difficulties of abstract thinking and they can cause the decrease of programming self-concept and other learning motives.

We redesigned the traditional learning environment. As a constructive approach project method was used. Our students worked in groups of two or three; small problems were solved after every lesson. In the problem solving process students use programmable robots (e.g. Surveyor, LEGO NXT and RCX). They had to plan their program, solve some technical problems and test their solution.

The usability of mobile robots in the learning process and the short-term efficiency of our teaching method were checked with a control group after a semester (n = 149). We examined the effects on our students' programming skills and on their motives, mainly on their attitudes and programming self-concept. After a two-year-long period we could measure some positive long-term effects.

*Keywords—programmable mobile robots; project method; positive effects*

## I. INTRODUCTION

Programming is not a compulsory subject in IT courses for students at Hungarian high schools. The National Curriculum aims at developing the skills of writing algorithm and developing algorithmic thinking but the skills of programming are taught only in classes that prepare students for IT graduation at advanced level. Because of this, many IT students start to acquire the elements of programming and program planning only during their college studies. Some previous research results [1] proved that those students who had already learned programming at high school were much more successful in the programming courses at our college. This advantage does not depend on the weekly number of high school lessons.

In contrast, beginners usually cannot pass their first exams. Lecturers often notice a decreasing increasing interest in programming. We supposed that abstract thinking means a great problem for beginners.

The Hungarian empirical research results [1] are supported by some results from other countries. In their comprehensive study on the Greek secondary school system Sartatzemi at al. [2] paid attention to the problems of teaching programming. They emphasize, there are some essential difficulties for those who have just started to learn programming:

- The professional programming languages are too complicated for beginners, in spite of the fact that these languages provide a wide range of solutions. Students usually have to focus rather on the language than on the problem itself. Accordingly, the implementation of a simple algorithm demands high-level thinking abilities.

- Professional programming environment is usually more complex than it is necessary for beginners. The environments do not help a beginner with the identification of syntax errors. The error messages in professional environments are made for professional programmers, not for beginners. The complexity of the environment can be shocking for students.

- During the first semester students cannot solve interesting problems. In order to enable them, they have to learn not only the programming language but the methods of developing larger programs as well. It is not possible during one semester. The grounding often seems too hard and boring for beginners and can decrease their motivation.

To sum it up, students have to focus not only on algorithm. They meet the principles of programming, the structure and syntax of the language, machine control problems etc. In addition, they have to learn the methods of program planning, developing and debugging.

The results of Sartatzemi et al. confirmed that new devices and methods are necessary in order to make the learning process more effective. Researchers usually propose different approaches so that beginners could cope with programming difficulties and with the complexity of programming languages. Some of them suggest that the object-oriented paradigm is more usable in teaching programming than the functional paradigm [3]. However, this change does not give a solution for the above-mentioned problems. Other researchers prefer a possible "learning programming language" [4] with an optimal environment and strongly limited set of statements.

We wanted to introduce new devices and methods into the learning process of programming. We looked for a method to improve the participation of students and increase communication among them. At the same time, we wanted to make devices more tangible. We aimed at making the learning process more concrete, practical and interesting for our students. The programmable mobile robot Mindstorms RCX (made by LEGO) appeared suitable for the realization of our aims. These devices and their programming environment allow students to learn in a natural, experimental way. Abstract thinking can be preceded by the manipulative and concrete usage of skills [5]. It can facilitate the development of skills and can deepen the level of understanding [6]. At the same time, the students' motives can increase due to the success of these learning situations. The co-operation among students can strengthen students' social and communicative skills [7]. These experience-based learning situations can lead to experiencing the growth of knowledge and can result in a higher level of students' activity. In a well-organized learning situation students can feel the flow. This is a mental state in which the students are fully immersed in concentration and the enjoyment of the activity [8]. These very motivated periods may accelerate skills development and may increase the efficiency of learning.

Our students' learning performance and the level of the adaptability of their knowledge are influenced by many factors. The effects of some factors have been revealed by researchers. The previously acquired levels of knowledge and skills have naturally a significant effect on the knowledge to be mastered. However, the individual differences in prior knowledge are not merely sufficient to explain the differences in further learning performances [9]. Additional factors have an important role in the learning process. The individual level of learning motivation (with many sub-factors) as well as family background or the system of social relations may have an effect on the students' learning success.

## II. REAL AND VIRTUAL ROBOTS IN EDUCATION

### A. Overview

One of the most interesting and most difficult problems in the field of artificial intelligence is to create and apply intelligent robots. Real robots have to work in a noisy, nondeterministic, continuous space and time it makes its necessary to solve a number of additional difficulties. Thanks to the burgeoning of low-cost high-performance computers, we are able to simulate robots in a virtual space. Working with robotic simulators programmers can focus on the algorithm, neglecting many of the real world's aspects.

In education, the question arises whether the use of real robots or the use of robotic simulators is more efficient in the development of students' programming skills. Using simulators the teacher can create and change the teaching environment. The complexity, inventiveness and realism of the environment can be adapted to the students' skills and age. However, students cannot "touch" the robots simulated on the screen. Because of this, the manipulative skill may be incomplete; this can cause difficulties in the process of interiorization [5].

Until the last decade robot simulators could be usually found in industrial applications. Additionally, some robot-specific simulators were used. In the last decade, possibilities of robot simulators moved towards general usability. By a special plug-in of MathLab, we can simulate punctiform robots or robots with real expansion [10].

An advanced simulation environment is the Webots mobile robotics simulator which is a commercial product developed by Cyberbotics[11]. It can simulate rolling, walking or flying robots. Additionally, this simulator can control some types of real robots (e.g. Pioneer, LEGO Mindstorms, Aibo) [12]. Repast (developed by the University of Chicago) is an open source, multi-agent simulation package based on Java [13] . The basic concepts of Repast were borrowed from the simulation environment of Swarm agent [14]. These simulators are used mainly in research.

In the past half a century some famous robot simulators were used in education too, e.g. Papert's turtle [15], Karel the robot [16] or the Spider World, used by Dalbey and Linn [17]. The Lego Mindstorms Simulator (LMS) developed by the University of Paderborn was also very popular in education. Some empirical experiences of teaching with robot simulators are shown in the next section.

### B. Research result

Sartatzemi et al. [2] used Mindstorms RCX mobile robots and ROBOLAB as a programming environment. In a ten-hour course (two hours/day), 14 students solved simple tasks in an icon-oriented environment. The teacher presented ROBOLAB structures in the first part of the lessons then the students solved tasks on worksheets. Researchers concluded that Mindstorms robots and the new programming environment are an efficient and practical way for high school students to learn programming. Their empirical data and the assessment showed some important conclusions. Students can easily acquire knowledge about procedures and the controlling of robots but this knowledge is often incomplete and inaccurate. The application of a real system is useful to analyze and solve a problem. Additionally, the students can check and debug their solutions in an experiential and clear way. It seems that students can understand the basic principles of programming in this environment. However, researchers observed some difficulties. The internal difficulties of the structures of the programs are similar to other environments; they can lead to misunderstanding. Because of this, the development of larger programs seems difficult for students. Furthermore, it was difficult for many students to connect the behavior of a robot to the logic of the program.

In their experiment, Wu et al. [18] compared the effectiveness of teaching with real and simulated robots. One of their groups (75 students) used LEGO RCX or LMR robots, the other group (76 students) used LEGO Mindstorms Simulators. Both groups consisted of beginners in programming. Similar pre-knowledge was supposed, so researchers did not use any pre-test. It was a short-term experiment (seven weeks; two hours/week); because of this researchers assessed the short time effects only. Pre-written templates in leJos (Java) language with simple program structures, basic variables and functions for controlling motors, lighting and crash sensors were used for the tasks. As their empirical result showed, there was not any significant difference between the two groups in understanding the pre-written programs and programming. However, those students who used real robots, showed a more positive attitude towards learning. Students in this group could easily imagine the behavior of robots but the usage of real robots demands extra time.

Kamada et al. [19] assumed that computer-controlled machines are used in almost all areas of life. Because of this, it is advisable to learn about the mechanism and the controlling opportunities of robots at the same time. Researchers assumed that the simultaneous learning of hardware and software can lead to an easier recognition and correction of errors of computer-controlled devices. Students built their own Myurobo robots, after that they tried to control them using Dolittle programming language. The simple and cheap robots could be used not only at school but at home as well. During the process of building the robots students acquired knowledge in the field of mechanics. The structure of the robots was similar to a "glass box", so the wiring of mechanical and electrical parts could be seen. According to the teachers' opinion, this way, the mechanical and electrical structure is easier to understand for the students. The experiment was organized as a pilot-project for high school students. The researchers did not aim to assess the effectiveness of learning; they were interested in whether these devices can be applied in education. The project lasted only 10 hours: two hours to learn the programming language, four hours to build the robots, four hours to solve a programming problem. The teachers wanted to find a programming language which can be learnt easily by the students. The low price (€20) and the "glass box" style of the device were attractive for the participants. An additional benefit was the simple control language. However, the environment had some disadvantages: the difficulty of serial-to-USB conversion and the fact, that there was no real-time control from the computer. The transformed control language also showed some differences from the original Dolittle language. Based on the feedback from the participants, the researchers considered that a revised, object-oriented Dolittle language can be applicable in high school education.

Kurebayashi et al. [20] prepared a proposal for a new curriculum for primary and secondary school students. They suggested introducing a new practice-oriented subject with tri-axial robots. With these robots, the elements of mechanics, electronics and information technology may be taught in one context. They examined whether the complex way of teaching with embedded systems is more effective than the traditional methods of teaching. The new curriculum was tested on a sample with 123 high school students. Students built robots and prepared their programs. After that a competition was organized for the robots. The effectiveness of the curriculum was measured by questionnaires for students and teachers. Based on the educators' feedback, the new curriculum resulted in positive effects and high efficiency. As the students looked back, building and programming a robot was a hard but enjoyable task. Many of them planned to continue learning about robot programming; this new curriculum sparked their interest in complex, systematic learning. To sum it up, the curriculum and its content may help with the teaching of complex embedded systems.

Fagin and Merkle [21] investigated the advantages of using robots in teaching information technology. They organized a control group experiment with more than 800 students and observed them for a year. It was expected that robot assisted learning would encourage students to choose computer engineering, computer science or any related field during their college studies. Additionally, researchers supposed that the robot can be a motivational device for students. Furthermore, they expected that programming skills would develop more significantly in the experimental group than in the control group. In the experimental group, the students used LEGO Mindstorms robots and Ada/Mindstorms as a programming environment. An additional aim was the acquisition of elements of the Ada language. In comparison to the above mentioned research results this experiment showed negative effects on the programming skills. The performance of students with the robots was significantly lower than in the control group taught with traditional devices and method. There are several possible reasons for this. After uploading, students had to compile and debug their programs on the robots; for this, more time was necessary compared to the traditional method using only computers. Another reason for errors may have been that teachers were well prepared for the lessons, but they had also used the robots for the first time, so they did not have sufficient experience in organizing robot assisted lessons. As the researchers summarized, despite their potential positive effects, the robots are not panaceas in education.

## III. Short-Term Effects of Using Project Method and Model Robots at Our College

### A. New course: new method and new devices

Because of problems described in the introduction, a new course was developed at our college. Our students learnt this course in a non-traditional way. LEGO NXT, LEGO RCX and Surveyor as programmable model robots were used to teach the elements of programming for IT-engineering students [22]. This new course can be taken by students who have successfully passed their "Programming 1." exam in C/C++ language. That is why NQC and NXC programming languages were chosen for this course. Syntax, statements, functions etc. of these languages are very similar to those used in standard C language. We did not put emphasis on the knowledge of the internal structure of robots. Our aim was to deepen our students' programming skills and algorithmic thinking, as well as to improve their attitude towards programming with these tangible devices.

NQC and NXC languages also contain loops, conditional statements, functions, tasks and included files similarly to standard C. From an educational point of view, one of the most important features is an easy way to run our program: we can upload it to the robot via Bluetooth, and check it immediately and visually.

In this course we rarely used traditional teaching methods e.g. teacher's presentation, but we often used methods giving an opportunity for constructive learning. The most preferred one was project method. Similarly to the business sector, in a project process the analysis of the problem, planning the steps towards their own solution and the implementation are carried out in groups [23]. The rigid and commanding knowledge transfer function of teachers has changed. Primarily, their roles are to raise the problem, to provide sufficient resources for work and to co-ordinate students' work. Simultaneously, the importance of their preparatory role has grown. Group members plan the process, divide the tasks among themselves, communicate to each other and, at the end of the work, they jointly summarize and present their results.

During this process, students could acquire theoretical and practical knowledge. This knowledge may be applicable in our students' future IT-engineering job and in their software developing projects as well. Teamwork can have positive effects on their communication skills because their thoughts, their ideas have to be expressed understandably but in a professional way [24]. Most courses of higher education rarely give opportunities for professional communication among students because of the high number of students.

In details, in our "Model robot programming" courses we aimed at following the principles, methods and processes of the constructive approach of teaching. Only 20% of time was used for teacher's presentation and explanation. During this period, the teacher introduced the subject, the necessary functions and the elements of the language via examples. In the remaining time, groups of two or three participants solved programming problems. The co-operation among the teammates was an important factor because they had to recognize the problem, to find a possible solution and to divide the job. As a teacher, we did not play a traditional knowledge distributor role in this phase. Instead of this, we had to support, motivate and co-ordinate our groups' work. We could help with identifying the main points of the problem, with accessing useful resources and sample libraries, with the accomplishment of the independent research etc. In the most significant period of the learning process the groups had to construct their robots and build them from LEGO Mindstorms or Surveyor parts. In this period they had to make and check their algorithm, write and debug their program. Additionally, a documentation of their solution with their plan, photos, video clips had to be prepared. At the end of the project each group presented their solution to the other groups and answered their questions. During the evaluation of the project, in addition to the teacher's reflections, self-evaluation, the other teams' and the teammates' evaluation also play an important role.

Our courses provided opportunities for collaborative knowledge building [25]. In a process like this, the understanding and interpretation of problems can be strengthened. The individual's activities for personal understanding are associated with social knowledge building [26].

### B. Short-term effects

With the new course introduced in the previous chapter, we wished to decrease the problems mentioned at the introduction. Based on the college course system, it was not possible to conduct an experiment for more than half a year. That is why we decided to monitor our students' results later in order to demonstrate the effectiveness of the development of motivation.

We presupposed that the usage of tangible devices may accomplish the activation and improvement of learning motives and the acquisition of basic elements of programming simultaneously [27].

H1: Real tools make learning more enjoyable.

The feeling of knowledge growth and joyful learning may lead to the flow state. It can work as a very strong learning motive. In addition, the gradually more complicated tasks may ensure a lasting strength of challenge. In this situation, the mastery motive can be activated and may play a fundamental role in skill acquisition.

H2: Programming self-concept can be improved with the use of robots.

The experiences obtained in robot programming and the achievements in problem solving tasks have an effect on students' self-confidence.

H3: The tasks solved by students at the concrete operational level have an impact on the development of abstract programming skills.

Acquiring programming at the abstract operation level often proves to be too difficult for starting programmers. We supposed that learning with tangible devices can enhance the acquisition of the abstract knowledge elements.

### 1) Methods

To verify the hypotheses, we organized a study with experimental and control groups. All students in the study took our course "Programming 1.". During the semester of our experiment, members of the experimental group (n1 = 73) used LEGO NXT robots with the methods introduced in chapter 3.1. Members of the control group were taught by traditional teaching methods.

We used a test with 15 items to assess our students' programming skills and knowledge (Cronbach-$\alpha$ = 0.86). Most Mitems required a short answer. In these items students had to understand short pieces of a program, after that they had to complete or debug them. We used the same test means for the pre-test and post-test.

In order to assess our students' programming self-concept and attitudes towards programming, we used a questionnaire containing 17 questions. To the majority of questions students could choose their answers from a five-level Likert-style response list. Six questions used for assessing the programming self-concept, were arranged into one factor (KMO = 0.87). We aggregated these variables into one new variable without weighting. This new variable was rescaled on a percent-point scale. The questionnaire contained some additional questions about students' social background.

Some more questions were asked in the post-test questionnaire. These questions concerned the hardness and fun of the work during the experimental semester.

The result of our two sub-samples in course "Programming 1." was very similar ($\chi 2$ = 3.86; p = 0.38). The pre-test difference between the experimental group and control group was not significant in their programming pre-knowledge and in their programming self-concept (Table 1).

There was a small, non-significant difference between the two sub-samples in the number of programming courses at high school (2 = 5.42; p = 0.27). Nearly half of the whole sample (46% of students) had not learned programming at high school.

TABLE I.    PROGRAMMING PRE-KNOWLEDGE AND PROGRAMMING SELF-CONCEPT IN THE SUB-SAMPLES IN THE PRE-TEST

|  | Experimental mean (st. dev.) | Control mean (st. dev.) | t (p) |
|---|---|---|---|
| Programming pre-knowledge (%p) | 44.6 (19.1) | 41.3 (19.5) | 1.36 (0.17) |
| Programming self-concept (%p) | 47.2 (19.7) | 46.3 (21.5) | 1.57 (0.14) |

Remark: we used Levene-F to compare standard deviations. The difference between standard deviations was not significant.

Based on similar results of our experimental and control groups, we supposed the differences at the end of the experimental semester were due to educational effects. The next chapter presents these changes

*2) Development of the experimental and control group*

We could not observe any significant development of programming skill either in the experimental group (xpre = 44.6 %p; xpost = 47.9 %p; t = -1.23; p = 0.23) or in the control group (xpre = 41.3 %p; xpost = 43.6 %p; t = -1.01; p = 0.34). The Pearson-correlation between the pre-test and post-test results are similar in the two sub-samples (rexp = 0.63; rctrl = 0.62). These results showed that the learning process with tangible devices had not directly affected our students' knowledge.

However, we could measure an important and significant difference between the experimental and control groups in the field of learning motives. During the semester members of the experimental group were absent on significantly less occasions than the members of the control group. The learning process was much more enjoyable for the experimental group (on a five-grade scale: xexp = 3.47; xctrl = 2.96; t = 3.87; p < 0.01), and they felt the course less difficult (xexp = 3.07; xctrl = 3.35; t = 1.96; p = 0.03). Simultaneously, the students' attitude towards their teacher did not change significantly (xpre-exp = 4.03; xpost-exp = 4.06; xpre-ctrl = 4.03; xpost-ctrl = 4.12), so we can suppose that the changes in the students' motives are not consequences of their teacher's personality.

The average programming self-concept of the control group remained unchanged during the experimental semester (xpre = 46.3 %p; xpost = 44.1 %p; t = -0.45; p = 0.66). However, the results showed significant changes in the experimental group members' programming self-concept (xpre = 47.2 %p; xpost = 52.2 %p; t = -2.60; p = 0.01). Differences between our sub-samples were also observed in the distributions of this variable.

These results showed that despite the short-period, significant changes in students' programming self-concept can be achieved using new devices and teaching methods. This is very important for the students' future learning performance because of the strong effect that well-developed self-concept has on learning achievement [28]. With monitoring students further we want to verify if the well-developed self-concept results in any additional programming effectiveness.

*C. Assessment of the durability of effects*

*1) Questions of our research*

As presented in chapter III.B, a short-term post-test showed positive effects on students' self-concept in the experimental group. However, we can check the durability of these effects and the usability of acquired knowledge only after a long-term period.

An important problem when organizing the long-term post-test was that most of our students in the sample had completed their college studies. That is why we could only involve our former students with available contact details in the assessment. The sample of the long-term post-test was limited by this fact. An additional problem was that it is almost impossible to organize a control group as it is very hard to create a sample whose members studied at the same time, whose previous measurement results and contact details are also available. So the assessment is based on the responses of students who studied our course previously.

Our analysis is primarily intended to clarify if the attitudes are long-lasting since the "Model robots programming" course was taken towards the topic and self-concept related to mobile robots programming. We also wanted to explore whether the beneficial short-term changes can be transferred to other areas of programming.

*2) Methods*

In the study 33 people took part. Previously, all of them had been involved in the course and the experiment introduced in chapter III.B. The total sample's average age is 30.9 years, standard deviation is 6.1 years. 36% of the sample was full-time, the others were correspondence students. Obviously, the full-time students' sub-sample (x = 26.2) was significantly younger than the correspondence students' (x = 33.6). The sample was considered to be a normal distribution of age. The sex ratio was not significantly different from what we can observe at the whole faculty, so we did not analyze the data in sub-samples of women and men. 15.2% of the sample was female.

We compiled a new questionnaire to assess long-term affects. The questions were related to completed studies as well as to other studies since then. We asked some questions about our former students' current job and its relationship to IT. The questionnaire consisted of 18 questions. These assessed the actual attitudes towards programming as well as to self-concept related to programming and mobile robot programming. Respondents could choose their answers from five-grade Likert-style lists.

An additional question was used for assessing our respondents' programming self-concept based on social comparison. They had to imagine a fictive situation where they had to fill in a 50-point programming test. Every respondent had to assess how many points he/she could collect if the average performance of his/her team mates was 35 points. So they had to give a norm-oriented assessment. This question could measure the respondents' self-concept [9] .

The questionnaire was sent to all former students of our course. They could answer the questions electronically. The questionnaire was sent back by 62% of those students whose contact details were available.

*3) Results*

The average length of pre-college IT courses was 3.5 years in the sample but 20% of respondents learned IT for only one academic year. The mean was significantly higher in the sub-sample of full-time students (x = 4.8 years) than in the sub-sample of correspondents (x = 2.6 years). The primary cause of this difference was the different age of the two sub-samples. We could observe a strong, significant Spearman-correlation between the age and the number of pre-college years with pre-college IT learning (r = -0.60; p < 0.01).

In the fields of programming, the mean of pre-college years is 1.6. There is a moderate but significant Spearman-correlation between the number of pre-college years with IT learning and the number of pre-college years with learning programming (r = 0.55; p < 0.01). The correlation between the age and the number of pre-college years with learning programming is also negative but lower, so the older sub-sample spent relatively more time with programming. It can be caused by the changes in the curriculum or by the changes in the fields of interest.

Students had learned our course for 3-5 years before our long-term post-test, so we did not analyze whether the answers depended on this variable. Because of the short period after graduation, only a very small proportion (just two people) gained further qualifications. 42.6% of the respondents deal with programming in their current work.

At the end of the "Model robot programming" course, the average mark was 4.6 (on a five-grade scale in Hungarian schools). This mean was calculated based on the memories of our respondents. It is similar to the average of the official results of all students who had passed this course (x = 4.56; n = 127). However, this mean is significantly higher than usual at the end of "Programming 1." courses (usually it is around III.A), despite the fact that the programming skills to be acquired are very similar in both courses. This may partly be caused by the fact that the students, who registered for this mobile robot programming course, had better pre-knowledge and motivation within the population. But in our sample the mean of "Programming 1." course mark was only 3.27, that is why we suppose that the difference is due to our experiment.

During further analysis we had to emphasize that quite a long time had passed since learning our course. So the long-term post-test could only analyze the durability of the long-term effects of attitudes and motives.

The "Model robot programming" course was considered easier (f = 60.6%) or much easier (f = 36.4%) by the respondents than other programming subjects. Only one respondent answered that this subject was more difficult for him and no one answered that it was much more difficult. There is no difference in this question between the full-time and the correspondence students (F = 1.97; p = 0.17; t < 0.01; p > 0.99). The feeling of difficulty was almost independent of age (r = 0.02; p = 0.94).

The durability of the respondents' positive attitude towards mobile robot programming was indicated by the great proportion of those (f = 90.9%) who found the subject much funnier and more enjoyable than other subjects. Only one ex-student remarked that it was as funny as any other programming subject. The difference between full-time students and correspondence students was not significant in this variable (Welch-d = 1.77; p = 0.10) and it is independent of age (r = 0.12; p = 0.63).

This positive attitude may be caused by many factors. One of them is, according to the respondents' opinion, that this course provided by far more possibilities for student activity than other course. This positive attitude was similar in the sub-samples of full-time and correspondence student (F = 0.03; p = 0.88; t < 0.01; p > 0.99) despite the fact that the number of contact lessons is much less for correspondence students.

Neither in the opinion regarding difficulty, nor in the attitudes towards mobile robot programming were any significant differences between the sub-samples of those who work in IT sector and those of working in other fields ("easier": F = 2.32; p = 0.14; t = 0.21; p = 0.83; "more enjoyable": F = 1.52; t = 0.23; t = 0.59; p = 0.56; "more possibility of activities": F = 8.51; p = 0.01; d = 1.79; p = 0.08). Based on these results, we supposed that the effect of "past became beautiful" was not significant in the case of those who work in IT now.

An indicator of programming self-concept may be a norm-oriented comparison of the students' own supposed result compared to his/her team's or classmates' supposed result in an imagined test (see Methods in this chapter). It is based on self-confidence and depends on self-evaluation compared to the peers' results. We measured this factor on a percent-point scale with a range of 0-100. The mean of the whole sample was 35.2 %p. It is a significantly better result for programming self-concept than what had been measured in earlier studies. In this study the distribution of this variable was very asymmetric, 65.6% of the sample gave a higher value than the reference value of earlier studies. However, there were some extremely low values so the distribution significantly differed from the normal distribution (Z = 1.65; p = 0,01). Assumed forgetting can be the reason for it but it also indicates that with these respondents positive self-image did not last long.

There was a significant difference in this variable between the sub-samples of full-time and correspondence students (xfull-time = 39.9; xcorr = 32.7; F = 11.40; p < 0.01; d = 2.19; p = 0.04). The individual differences were much bigger in the sub-sample of correspondence students. This result supported our earlier experiments: one of the reasons for the choice of correspondence courses – together with family and social backgrounds – is the lower learning self-concept.

Those who work in IT sector had a small advantage in this variable (xIT = 37.6; xothers = 33.3) but this difference is not significant (F = 0.51; p = 0.48; t = 1.07; p = 0.29). This small difference could be a result of further workplace successes.

The self-concept related to programming and to mobile robot programming were assessed by six-six Likert-style questions. Both groups of these variables were arranged into one factor (KMOprog = 0.84; KMOmobile = 0.72). Based on this we aggregated these variables into two new variables and transformed them into percent-point scale. Both of the new variables showed a normal distribution (programming self-concept: Z = 0.95; p = 0.33; self-concept related to mobile robot programming: Z = 0.45; p = 0.98).

There was a great and significant difference between these two variables in the whole sample (xprog = 56.7 %p; xmobile = 88.5 %p; t = 6.77; p < 0.01). The Pearson-correlation between two variables is not significant (r = 0.11). This result showed that these factors are independent of each other despite the fact that mobile robot programming is a sub-field of programming. The mean of programming self-concept was similar to the results of earlier studies. However, the self-concept related to mobile robot programming was correlated to the above mentioned factor of norm-oriented comparison (r = 0.44; p = 0.02), increasing the validity of our result.

The self-concept related to programming and to mobile robot programming was similar in the two sub-samples of full-time and correspondence students (programming self-concept: xfull-time = 55.9 %p; xcorr = 57.2 %p; self-concept related to mobile robot programming: xfull-time = 86.7 %p; xcorr = 89.5 %p).

IV. Conclusions

Our results indicate that the learning process with programmable mobile robots and with new teaching methods could improve the attitude towards mobile robot programming and self-concept, however, we could not observe any significant transfer effects to other fields of programming. This fact was underpinned by the negative and significant Spearman-correlation between the programming self-concept and the marks at the end of the "Model robot programming" course (r = -0.47; p < 0.01). The successful transfer of the mobile robot programming self-concept to other programming areas would need further positive results.

Acknowledgment

This research was supported by the European Union and the State of Hungary, co-financed by the European Social Fund within the framework of TÁMOP 4.2.4. A/2-11-1-2012-0001 'National Excellence Program'